\documentclass[aps,prd,showpacs]{revtex4}  
\usepackage{amsmath}
\usepackage{graphicx}

\def\eq#1{{Eq.~(\ref{#1})}}

\newcommand{\LL}{Lanczos-Lovelock}
\newcommand{\LDm}{\ensuremath{L_{(m)}}}
\newcommand{\Riem}[4]{\ensuremath{R^{#1 #2}_{#3 #4}}}
\newcommand{\sD}[1]{\sum_{m=1}^{K}{#1}}
 
\newcommand{\Alt}[6]{\ensuremath{\delta^{#1 #2 ... #3}_{#4 #5
      ... #6}}}

\begin{document}
\title{Structure of \LL\ Lagrangians in Critical Dimensions}
\date{August 29, 2010}

\author{Alexandre Yale}
\email{ayale@perimeterinstitute.ca}
\affiliation{University of Waterloo, 200 University Avenue West, Waterloo, Ontario N2L 3G1, Canada}
\affiliation{Perimeter Institute, 31 Caroline St. N., Waterloo, Ontario N2L 2Y5, Canada}
\author{T.  Padmanabhan}
\email{paddy@iucaa.ernet.in}
\affiliation{IUCAA, Post Bag 4, Ganeshkhind, Pune - 411 007, India}

\begin{abstract}
The \LL\ models of gravity constitute the most general theories of gravity in $D-$dimensions which satisfy (a) the principle of of equivalence, (b) the principle of general covariance, and (c) have field equations involving derivatives of the metric tensor only up to second order. The $m$th order \LL\ Lagrangian is a polynomial of degree $m$ in the curvature tensor. The field equations resulting from it become trivial in the critical dimension $D= 2m$ and the action itself  can be written as the integral of an exterior derivative of an expression involving the vierbeins, in the differential form language. While these results are well known, there is some controversy in the literature as to whether the \LL\ Lagrangian itself can be  expressed as a  total divergence of quantities built only from the metric and its derivatives (without using the vierbeins) in $D=2m$. We settle this issue by showing that this is indeed possible and provide an algorithm for its construction. In particular, we demonstrate that, in two dimensions, $R \sqrt{-g} = \partial_j R^j$ for a doublet of functions $R^j=(R^0,R^1)$ which depends only on the metric and its first derivatives.  We explicitly construct families of such $R^j$-s in two dimensions. We also address related questions regarding the Gauss-Bonnet Lagrangian  in $D=4$. Finally, we demonstrate the relation between the Chern-Simons form and the $m$th order \LL\ Lagrangian.
\end{abstract}


\maketitle

\newcommand{\myc}[2]{\left( \begin{array}{c} #1 \\ #2 \end{array} \right)}

\newcommand{\myeq}[2]{\begin{equation} \begin{split} \label{#1} #2 \end{split} \end{equation}}

\section{Introduction and motivation}\label{sec:intro}
Consider a theory of gravity in $D$ dimensions  based on the action 
\begin{equation}
A=\int \sqrt{-g}\ d^Dx [L(R^a_{\phantom{a}bcd}, g^{ab})+L_{\text{matt}}(g^{ab},\phi_A)],
\end{equation} 
 where $L_{\text{matt}}(g^{ab},\phi_A)$ is a suitable matter Lagrangian involving some matter degrees of freedom $\phi_A$. Varying $g^{ab}$ in this action, with suitable boundary conditions,  leads to the field equations (see e.g. sec 3.5 of ref. \cite{rop} or, for a textbook description, chapter 15 of ref.\cite{gravitation}):
 \begin{equation}
\mathcal{G}_{ab}=P_a^{\phantom{a} cde} R_{bcde}  - \frac{1}{2} L g_{ab}- 2 \nabla^c \nabla^d P_{acdb}
\equiv \mathcal{R}_{ab}-\frac{1}{2} L g_{ab}- 2 \nabla^c \nabla^d P_{acdb}=\frac{1}{2}T_{ab},
\label{genEab}
\end{equation}
where
\begin{equation}
P^{abcd} \equiv \frac{\partial L}{\partial R_{abcd}}.
\end{equation} 
We use a signature $(-,+,+,...)$  and units with $G_N = c=1$; Latin indices run over $(0,1,2,...., D-1)$.
The notation in terms of calligraphic fonts is suggested by the fact that in Einstein's theory, with $L= R/16\pi$, we have $\mathcal{R}_{ab}=R_{ab}/16\pi$ and $\mathcal{G}_{ab}=G_{ab}/16\pi$ with standard normalization. The field equation in \eq{genEab} will not contain terms involving derivatives of the metric of degree greater than two if $\nabla_a P^{abcd} =0$. Since $P^{abcd}$ has the symmetries of the curvature tensor, it follows that it must be divergence-free in all indices. One can show that the most general tensor satisfying such conditions, built from the metric and curvature tensor, must be a polynomial with the indices of curvature tensors contracted out in a specific manner using the determinant tensor. In such a case, the gravitational part of the Lagrangian will take the form
\begin{equation}
{L} = \sD{c_m\LDm}\,~;~{L}_{(m)} \propto \Alt{a_1}{a_2}{a_{2m}}{b_1}{b_2}{b_{2m}}
\Riem{b_1}{b_2}{a_1}{a_2} \cdots \Riem{b_{2m-1}}{b_{2m}}{a_{2m-1}}{a_{2m}}
\,,  
\label{twotw}
\end{equation}
where \Alt{a_1}{a_2}{a_{2m}}{b_1}{b_2}{b_{2m}} is the determinant tensor (for example, $\delta^{ab}_{cd} = \frac{1}{2}(\delta^a_c \delta^b_d - \delta^a_d\delta^b_c)$), the $c_m$ are arbitrary constants and \LDm\ is called the $m$-th
order \LL\ Lagrangian \cite{Lanczos1932}.
The $m=1$ term is proportional to $\delta^{ab}_{cd}R^{cd}_{ab} \propto R$ and leads
to Einstein's theory. The $m=2$ term gives rise to what is known as Gauss-Bonnet theory
with
\begin{equation}
L_{\text{GB}} \equiv L_2\propto(R^2 - 4R_{ab}R^{ab} + R_{abcd}R^{abcd}).
\label{gbl}
\end{equation} 
 In the language of differential forms and tetrads, denoting our cotangent basis $\omega^a$, spin connection $\Omega^{AB}$, and curvature two-form $\Theta^{AB}$, and defining $\epsilon_{a_1 \cdots a_n} = *(\omega_{a_1} \wedge \cdots \wedge \omega_{a_n})$, the $m$th \LL\ Lagrangian form $\mathcal{L}_m$ can be defined as:
\myeq{defineLm}{
\mathcal{L}_m & \equiv \Theta^{A_1 B_1} \wedge \cdots \wedge \Theta^{A_m B_m} \wedge \epsilon_{A_1 B_1 \cdots A_m B_m}. 
}
(See, for example, \cite{PhysRevD.36.392}; we will try to consistently use calligraphic font for the differential form corresponding to the Lagrangian density and normal font for Lagrangian which is scalar; in most case, the context will make clear what is the expression we are talking about). \LL\ Lagrangians are the most natural generalizations of the Einstein-Hilbert Lagrangian to higher dimensions, and  both share several remarkable properties. It has been  shown \cite{Mukhopadhyay:2006vu} that these Lagrangians embody a version of the holographic principle, at the level of the action, and the resulting models share many of the thermodynamic features of the standard General Relativity \cite{Kothawala:2009kc,LLthermo}.

Because of the determinant tensor in \eq{twotw}, it is obvious that in any given dimension $D$ we can only have $K$ terms where 
$D\geq 2K$.
We will call $D=2m$ the \textit{critical dimension} for the $m$th order  \LL\ Lagrangian.
It is possible to show by straightforward combinatorics (see e.g. \cite{Kothawala:2009kc}) that $\mathcal{G}_{ab} =0$ identically in the critical dimension; this generalizes the well known result that $G_{ab}=0$ identically in $D=2$, which is the critical dimension for Einstein-Hilbert action corresponding to the $m=1$ \LL\ Lagrangian. It follows that  the field equations are vacuous for the \LL\ theory at the critical dimension.

This raises the question: What is the structure of \LL\ \textit{action itself} in the critical dimension? The vacuous nature of the field equations tells us that the \textit{variation} of the action will be a pure surface term in the critical dimension and can be ignored (or compensated) under suitable boundary conditions. But it is not obvious whether the action itself is a pure surface term. This is the central question we address in this paper. We begin with a series of comments  to convince the reader why this issue is interesting and nontrivial.

Let us focus our attention on $D=2$, which is the critical dimension for $m=1$ \LL\ theory, the Lagrangian of which is simply $R$ and the action is just the standard Einstein-Hilbert action. It is trivial to show, using the language of differential forms, that the integrand of the Einstein-Hilbert  action $R \sqrt{-g} $  in $D=2$ can be written as an exterior derivative of a quantity which explicitly involves the \textit{tetrads} (see \eq{R=dOmega} below). But what we would like to know is whether the integrand of the Einstein-Hilbert  action $R \sqrt{-g} $ can be expressed as $\partial_j R^j$ for a doublet of functions ($R^0, R^1$) built \textit{only} from the metric and its derivatives \textit{without} using tetrads in a \textit{general} coordinate system. That is, if we take the line element to be 
\begin{equation}
 ds^2 = A (t,x) dt^2 +2 C(t,x) dx dt + B(t,x)dx^2,
\end{equation} 
(in which we have not imposed any gauge or coordinate conditions) we want an expression for  ($R^0, R^1$) in terms of  $A, B, C$ and their derivatives. This is again trivial to accomplish if we use the conformally flat gauge with $C=0,A=B$; then one gets $R\sqrt{-g}=\partial_j\partial^j \ln A$ which has the required form but this will \textit{not} survive when we transform to an arbitrary coordinate system. 

Given the fact that the answer in the language of differential forms is already known, one would have thought that the expression for $R^j$ would already exist in the literature. We found, to our surprise, that this is not the case! Further,  a paper by Deser and Jackiw \cite{Deser:1995ne} seems to suggest that this is impossible, which had led Kiriushcheva and Kuzmin \cite{Kiriushcheva:2006nn} to recently conclude that there were fundamental differences between the vierbein formalism and the standard formalism in General Relativity (most notably, the vierbein formalism is known to have an extra internal gauge symmetry) and, because of those differences, the Einstein-Hilbert Lagrangian being a total derivative in the vierbein formalism does not imply that it is a total derivative in the standard formalism.  
It is therefore clear that this issue deserves clarification: on one hand, several papers claim (at least implicitly) that the action itself is a total derivative for all \LL\ models in critical dimensions but without concrete proof; on the other hand, there are arguments,  summarized in the articles by Kiriushcheva and Kuzmin \cite{Kiriushcheva:2005kk,Kiriushcheva:2006nn}, that these claims are ill-founded.  This confusion is in large part due to the lack of an explicit expression for $R^j$, as a function of the metric and its derivatives, satisfying $L_m \sqrt{-g} = \partial_j R^j$.  One of our conclusions is that $R^j$  can indeed be expressed in terms of the metric and its derivatives. We will provide an explicit expression for $R^j$ thereby putting an end to the debate in the case of $D=2$.
This $R^j$ is highly non-unique: we can always add to it any set of functions $f^j(x)$ which satisfy $\partial_jf^j=0$. We will see that there is one natural $R^i$ associated with each unit normalized vector field in the $D=2$ spacetime and, of course, any two of them differ by an $f^j$ with  $\partial_jf^j=0$.

Let us next turn to  the case of $D=4$. The \LL\ Lagrangian which becomes critical in $D=4$ is the $m=2$ Gauss-Bonnet term, given by \eq{gbl}. The question again is whether one can write $L_{\rm GB} \sqrt{-g} = \partial_j R^j$.
Such a claim, with an expression for $R^j$, has been given in \cite{Cherubini:2003nj} and has been cited in a few later papers as well based on an expression originally given in \cite{precherubini}. Unfortunately, this expression is incorrect and arose from an invalid identification of frame indices with spacetime indices.
 We will clarify this issue as well and will provide a proof that every critical-dimension \LL\  Lagrangian can be written as a total derivative of functions of the metric and its derivatives, and that these functions can be thought of as the coordinate and local representation of the familiar Chern-Simons form.  This will be done in a pedagogical manner to ensure that the issues are clarified.

We will now proceed to establishing these claims. The plan of the paper is as follows:  we begin by addressing the issues related to the two-dimensional Einstein-Hilbert Lagrangian in Section \ref{sec:2DEH} by explicitly finding the pair of functions $R^j(g,\partial g)$ which satisfy $R \sqrt{-g} = \partial_j R^j$.  We then briefly consider the four-dimensional Gauss-Bonnet Lagrangian in Section \ref{sec:4DGB} before moving on to the general $2m$-dimensional \LL\ Lagrangians in Section \ref{sec:2mDLL}.  Appendix \ref{sec:CartanReview} gives a short review of the Cartan formalism which is used throughout the paper, while Appendices \ref{app:der1} through \ref{chern} present in  detail the derivations to some of our claims.

\section{ Einstein-Hilbert action in $D=2$} \label{sec:2DEH}
We begin by considering the two-dimensional Einstein-Hilbert Lagrangian and will obtain an explicit  expression for $R^j$, dependent only on the metric and its derivatives, such that $R \sqrt{-g} = \partial_j R^j$.  This will provide a direct proof that the Lagrangian is indeed a total derivative.  We will do so in two ways to provide two  perspectives on how the form of $R^j$  could have been obtained: the first procedure derives $R^j$ through the Cartan formalism which uses the language of differential forms. This is quite powerful and immediately generalizes to arbitrary $D$.
The second procedure, specific to $D=2$,  constructs one $R^j$ from each unit normalized vector field.  Finally, we mention a procedure based on conformal transformation, leaving the details to Appendix \ref{app:der2}.
These various approaches are discussed because there has been some controversy regarding differences between the Cartan formalism and the standard formalism; hence, it is useful to derive \eq{FG} without resorting to the Cartan formalism at all.  

\subsection{\label{der1} First Method: Cartan Formalism}
If we define an orthonormal basis $\omega^a = \omega^a_i dx^i$ satisfying $g_{ij} = \omega^a_i \omega^b_j \eta_{ab}$ as a basis for our cotangent space, then we notice that the spin connection, defined by the first Cartan structural equation (see \eq{cartan1} of Appendix A) must be antisymmetric: $\Omega_{01} = -\Omega_{10}$.  Then, the Einstein-Hilbert two-form can be written (up to an overall sign due to the definition of $d^2x$):
\myeq{R=dOmega}{
R \sqrt{-g} d^2x &=  \Theta^{ab} \wedge * (\omega_a \wedge \omega_b) \\
&= d \Omega_{01} d^2x,
}
where $\Theta^{ab} = \left( d \Omega^{ab} + \Omega^a_{~s} \wedge \Omega^{sb} \right)$ is the curvature two-form (see Appendix B for details).  Therefore, finding $R^j$ such that $R \sqrt{-g} = \partial_j R^j$ holds is straightforward: one simply needs to calculate explicitly the spin connection $\Omega_{01}$ and take its exterior derivative.  This is done in Appendix \ref{app:der1}, and the result is
\begin{widetext}
\begin{equation} \begin{split} \label{FG}
R^0 &= \frac{1}{\sqrt{-g}} \left[ -\lambda \frac{g_{01}}{g_{00}}g_{00,1} + (1-\lambda)\frac{g_{01}}{g_{11}}g_{11,1} - 2(1-\lambda)g_{01,1} + g_{11,0} \right] \\
R^1 &= \frac{1}{\sqrt{-g}} \left[ -(1-\lambda) \frac{g_{01}}{g_{11}}g_{11,0} + \lambda \frac{g_{01}}{g_{00}}g_{00,0} - 2\lambda g_{01,0} + g_{00,1} \right],
\end{split} \end{equation}
\end{widetext}
which is a one-parameter family of solutions, parametrized by a constant $\lambda$.  This parameter should be thought of as characterizing the gauge freedom we have in defining an orthonormal basis for our cotangent space. The $R^j$s for two different values of $\lambda$
constitute an example of non-uniqueness as mentioned in Sec. \ref{sec:intro}.  One can directly verify that the difference $f^j\equiv [R^j(\lambda_1)- R^j(\lambda_2)]$ satisfies $\partial_jf^j=0$, as it should. We could have, of course, set say $\lambda=1$ without any loss of generality using this freedom; we have retained for future convenience. (This corresponds to the freedom of changing $\Omega_{01}$ to $\Omega_{01}\to\Omega_{01}+f$ where $df=0$.)  It should also be emphasized that this gauge freedom is entirely unphysical, and one can fix it without imposing any condition on our spacetime or our metric.  In this sense, the gauge-fixed Cartan formalism is equivalent to the standard formalism for General Relativity.

\subsection{\label{der2} Second Method: $R^j$ from normalized vectors}
In two dimensions, we know that $R_{ab}=(1/2)g_{ab}R$; this will allow us to express $R$ as the total divergence of a vector field by the following procedure. Let $n^i$ be a unit normalized vector in two dimensions which we take to be timelike (with $n_in^i=-1$) for definiteness; the same procedure works for spacelike vectors as well.  Using the standard relation between $n^bR_{ab}n^a$ and the commutator of covariant derivatives, it is easy to show that
\begin{equation} 
R_{bd} n^b n^d = n^b \nabla_a \nabla_b n^a - n^b \nabla_b \nabla_a n^a 
= \nabla_a (n^b \nabla_b n^a - n^a \nabla_b n^b) + M^b_{\phantom{b}b}M^a_{\phantom{a}a}  - M^a_{\phantom{a}b} M^b_{\phantom{b}a},
\label{key1}
 \end{equation}
where we defined the tensor $M_{ab} \equiv \nabla_a n_b$. The last two terms cancel out in two dimensions.  To see this, note that $[M^b_{\phantom{b}b}M^a_{\phantom{a}a}  - M^a_{\phantom{a}b} M^b_{\phantom{b}a}]$ is a scalar; since $n^bM_{ab}=0$, then, choosing coordinates  where $n^a \propto \delta^a_0$, we find that $M_{a0}=0$ and $M^a_{\phantom{a}0}=0$. This then implies
\begin{equation} 
M^b_{\phantom{b}b}M^a_{\phantom{a}a}  - M^a_{\phantom{a}b} M^b_{\phantom{b}a}
= M^1_{\phantom{1}1}M^1_{\phantom{1}1}  - M^1_{\phantom{1}1} M^1_{\phantom{1}1}
= 0.
\label{key2}
\end{equation}
Further, using $R_{ab}=(1/2)g_{ab}R$ and $n_in^i=-1$, we combine \eq{key2} and \eq{key1} to find 
\begin{equation} 
R = 2\nabla_a (n^a \nabla_b n^b- n^b \nabla_b n^a) .
\label{Rfromn}
\end{equation}
It follows that $R\sqrt{-g}=\partial_aR^a$, where 
\begin{equation}
R^a=2\sqrt{-g}(n^a \nabla_b n^b- n^b \nabla_b n^a),
\label{key3}                                                  
\end{equation}  
for \textit{any} unit normalized vector $n^a$. Since we get one $R^a$ for each $n^a$, there is infinite degeneracy in the expression; any two such $R^j$ will differ by an $f^j$ with $\partial_jf^j=0$. 

The derivation above is almost identical to the standard Gauss-Codazzi decomposition (see e.g. (12.55) of \cite{gravitation}) of $R$ in terms of the extrinsic curvature $K_{ij}=-M_{ij}-n_i(n^k\nabla_kn_j)$. In two dimensions, the only non-zero component of $K^{ij}$ is $K^{11}$, which means that $Tr(K^2)=(TrK)^2$.  Since we also have $(TrK)^2=(TrM)^2$ and $Tr(K^2)=Tr(M^2)$, then this decomposition is identical to \eq{key2} and we get $R=-2 \nabla_j(Kn^j+a^j)$ with $a^j=n^b \nabla_b n^j$. We gave a direct derivation of \eq{Rfromn} to stress  that we need not introduce hypersurface orthogonality or a foliation to obtain the above result; the fact that $M_{ab}$ is not symmetric is also irrelevant. Of course, this method only works because we are in two dimensions.

By working out the components of $R^j$ with $n^i=(-g_{00})^{-1/2}\delta^i_0$, one obtains
\begin{equation} 
\begin{split}
R^0 &= \frac{1}{\sqrt{-g}} \left[ - \frac{g_{01}}{g_{00}} g_{00,1} + g_{11,0} \right] \\
R^1 &= \frac{1}{\sqrt{-g}} \left[ \frac{g_{01}}{g_{00}}g_{00,0} - 2g_{01,0} + g_{00,1} \right]
\end{split} 
\label{compo}
\end{equation}
which is the same as \eq{FG} with $\lambda=1$. Repeating the derivation with a spacelike vector $n^i\propto\delta^i_1$, one would obtain the corresponding result with $\lambda=0$.  The linear combination of these solutions $R^j$ is also a solution, giving us \eq{FG}.
It is therefore clear that $R \sqrt{-g}$ can indeed be written as the total derivative of a function of the metric and its derivatives in any given coordinate system in an infinitely degenerate way. 

In the paper by Deser and Jackiw \cite{Deser:1995ne} mentioned earlier, it is stated that ``....$R^\mu$ cannot be presented explicitly and locally in terms of the generic metric $g_{\mu\nu}$ and its derivatives $\partial_\alpha g_{\mu\nu}$ ....." . One of the authors has informed us (Jackiw, private communication) that they were not interested in an expression for $R^j$ given in terms of components. In that case, the sentence quoted above does not exclude expressions such as \eq{FG}, \eq{compo} etc. which use the components explicitly.  
In contrast, \eq{key3} does give an expression for $R^j$ explicitly as a vector density, but uses an external construct --- the vector field $n^i$ --- in addition to the metric and the connection (and hence again does not contradict the statement quoted above). While one cannot construct nontrivial tensorial quantities (except, of course, for the density $\sqrt{-g}$) from the metric and its first derivatives, it is certainly possible to do so if additional vector fields are given. 

To avoid misunderstanding we stress the following point: 
In \textit{any}  coordinate system $\left\{ x^i \right\}$, one can choose a normalized timelike vector  $n^i=(-g_{00})^{-1/2}\delta^i_0$ with components proportional to $(1,0)$ and use \eq{key3} to obtain \eq{compo} in that particular frame. 
If we now perform a coordinate transformation $x^i\to \bar x^i$, \eq{key3} will transform in a generally covariant manner with $R^j(\sqrt{-g})^{-1/2}$ transforming as a vector.  
However, since $\bar n^i$ will no longer be proportional to $(1,0)$, the explicit expression, \eq{compo}, will not hold in the new coordinate system. 
Instead, there will exist another vector,  $N^j=(-\bar g_{00})^{-1/2}\delta^i_0$, in the \textit{new} coordinate system which yields the same expression, \eq{compo}, in terms of $\bar g_{ab}$. (Recall that \eq{Rfromn} holds for any normalized timelike vector field; each gives one particular $R^j$ through \eq{compo}.)  Similar results hold for a unit normalized spacelike vector field. 
\textit{In other words, in any frame, we can always choose a vector $n^i$such that \eq{key3} reduces to \eq{compo}.}  In this sense, the expression in \eq{compo} holds \textit{in the same form} in all frames, though in each frame it is associated with a different vector field. Therefore, expressions like \eq{FG} are as good as expressions involving abstract indices (e.g. $\partial_jg_{ab}$) rather than explicit components (e.g. $\partial_0g_{01}$). 

It is also possible to obtain \eq{FG} by demanding that $R^j$ have the correct conformal transformation properties; the details of this method are in Appendix \ref{app:der2}.

\section{ Gauss-Bonnet Lagrangian in $D=4$} \label{sec:4DGB}
Before moving on to the general \LL\ action in arbitrary dimension, we shall quickly consider the four-dimensional Gauss-Bonnet Lagrangian.  We begin by rewriting $L_{GB}\sqrt{-g}$ in the differential form language:
\myeq{works1}{
 \mathcal{L}_{GB}&= (d \Omega^{AB} + \Omega^A_{~S} \wedge \Omega^{SB}) \wedge \\
&~~~~~~~(d \Omega^{CD} + \Omega^C_{~T} \wedge \Omega^{TD}) \wedge \epsilon_{ABCD} \\
&= d \left[ \Omega^A_{~B} \wedge \left( \Theta^C_D - \frac{1}{3} \Omega^C_{~F} \Omega^F_{~D} \right) \wedge \epsilon_{A~C}^{~B~D} \right], 
}
This equation is derived (see Appendix \ref{cherubini}) in the orthonormal basis where the spin connection $\Omega_{AB}$ is antisymmetric; it is the analogue of \eq{R=dOmega}. Proceeding exactly as in the case of $D=2$, we can, in principle, obtain an expression similar to \eq{FG} thereby expressing the $R^j$ in terms of the metric and its derivatives. In practice, however, this calculation, while conceptually simple, becomes tediously long due to the overwhelming number of terms involved.  The reason for this is easy to understand: in the $m=1$ case, we had $\mathcal{L}=d \Omega$ with only one non-zero $\Omega_{ab}$ term: $\Omega_{01}$.  This term was built from two function $\mu_{bc}^{~~~a}$ as per \eq{mu2} from Appendix \ref{sec:CartanReview}, each of which is built of up to two vierbein terms.  We therefore end up with approximately $1 \times 2 \times 2 = 4$ terms for $R^j$.

In the $m=2$ case, each $R^j$ is actually built from a sum of 36 terms, each of the form $\Omega \wedge (d \Omega + \Omega \wedge \Omega - \Omega \wedge \Omega)$.  Each such $\Omega$ is built from eight $\mu_{bc}^{~~~a}$ terms, each of which is built from four terms.  So at first estimate, we would end up with $36 \times (8 \times 4)^3 \approx 10^6$ terms.  While this is a rough calculation (and overestimate), it is clear that the expressions will be far larger than they were in the $D=2$ case.  Nevertheless, we will illustrate the procedure for one particular metric: consider the line element
\myeq{simplemetric}{
ds^2 = g_{00}(t)dt^2 + 2g_{01}(t)dtdx + 2g_{02}(t)dtdy + 2g_{03}(t)dtdz + g_{11}(t)dx^2 + g_{22}(t)dy^2 + g_{33}(t)dz^2,
}
where every component depends exclusively on the time coordinate.  Then, using the steps outlined in Appendix \ref{app:der1}, one finds
\myeq{R0GB}{
R^0 = &-g_{22,0}g_{11,0}g_{33,0}g_{00}g_{33}g_{22}g_{11} \times \\
&\frac{  -g_{11}^3g_{22}g_{00}^3 + g_{11}^3g_{02}^2g_{00}^2 + 3g_{11}^2g_{22}g_{01}^2g_{00}^2 - 2g_{11}^2g_{02}^2g_{01}^2g_{00} - 3g_{11}g_{01}^4g_{22}g_{00} + g_{11}g_{01}^4g_{02}^2+g_{01}^6g_{22} }{\left( -g_{00}g_{11}g_{22} + g_{22}g_{01}^2 + g_{02}^2g_{11}\right) ^{5/2} \left( -g_{00}g_{11}+g_{01}^2 \right)^2 \left( \frac{g_{00}g_{11}g_{22}g_{33} - g_{11}g_{03}^2g_{22} - g_{11}g_{33}g_{02}^2 - g_{33}g_{22}g_{01}^2}{g_{22}g_{00}g_{11} - g_{22}g_{01}^2 - g_{02}^2g_{11} } \right)^{3/2} (-g_{00}) },
}
which satisfies
\myeq{GBAnswer}{
\sqrt{-g} \left( R^2 - 4R_{ab}R^{ab}+R_{abcd}R^{abcd} \right) = \partial_0 R^0.
}
It would therefore be advantageous to find a way to write $R^j$ using abstract indices, Christoffel symbols, and other well-known quantities in order to make it easier to work with. 
Further note that --- because of the non uniqueness of $R^j$ --- it is entirely possible that the metric in \eq{simplemetric} admits another $R^j$, which is a lot simpler and differs from the one in \eq{R0GB} by a set of functions $f^j$ with $\partial_jf^j=0$.  It is very unwieldy to start with an expression such as \eq{R0GB} and attempt to rewrite it or simplify it.  However, as we shall see in the rest of this section and in Section \ref{sec:killing}, there are very large classes of spacetimes which do actually allow us to write $R^j$ in rather simple forms (In fact, as we will see in Section \ref{sec:killing}, there are indeed much simpler forms of $R^j$ available for the metric in \eq{simplemetric}).

Cherubini et al.\cite{Cherubini:2003nj}  has  an expression for $R^j$ for the four-dimensional Gauss-Bonnet Lagrangian which is based on \cite{precherubini}. This expression is the only one for $R^j$ (for any \LL\ Lagrangian) which we could find  in the literature:
\myeq{Cherubini}{
 L_{GB}\sqrt{-g} = -\partial_a\left[ \sqrt{-g} \eta^{a b c d} \eta_{r s}^{\phantom{rs}m n} \Gamma^r_{m b} \left( \frac{1}{2} R^s_{n c d} + \frac{1}{3} \Gamma^s_{c l} \Gamma^l_{n d} \right) \right],
}
where $\eta$ refers to the Levi-Civita symbol. We found its simplicity amazing, but we realized that --- unfortunately --- this equation is algebraically incorrect: it fails for just about any metric.  By a curiously incorrect reasoning, one can obtain a similar expression which works for a large class of metrics \cite{comment2}. We describe this result briefly, in view of its potential interest and to correct some of the errors in the formulas in the literature.

Note that our expression in \eq{works1} is derived in the orthonormal basis in which one cannot --- in general --- express the spin-connections in terms of Christoffel symbols by   $\Omega^a_{~b} = \Gamma^a_{bc} dx^c$. This relation, however,  is true in a coordinate basis (for a  review of the differential form language, see e.g., chapter 11 of ref.~\cite{gravitation} or Appendix \ref{sec:CartanReview}).  Interesting things happen if we \textit{forget} this fact and substitute $\Omega^a_{~b} = \Gamma^a_{bc} dx^c$ (which is certainly incorrect) for the spin connections in \eq{works1}. In that case, we find
\myeq{FixedCherubini}{
L_{GB}\sqrt{-g}&= -\partial_a\left[ \sqrt{-g} \eta^{a b c d} \eta_{r s}^{\phantom{rs}m n} \Gamma^r_{mb} \left(\frac{1}{2} R^s_{ncd}  -\frac{1}{3} \Gamma^s_{cl} \Gamma^l_{nd}  \right) \right]\equiv \partial_aQ^a.
}
The expressions in the right hand sides of \eq{Cherubini} and \eq{FixedCherubini} differ only by a sign of a $\Gamma^2$ term.  (This sign difference, it turns out, is simply due to confusion caused by differing conventions between \cite{precherubini} and \cite{Cherubini:2003nj}).
This expression in \eq{FixedCherubini} however --- quite surprisingly, since we made an obviously incorrect assumption --- happens to be correct for a large number of spacetimes, such as Schwarzschild, Kerr, Milne, Godel and even Ozvath-Schuking.  This expression also has the advantage of being written in a concise notation: it uses abstract indices and familiar quantities such as Christoffel symbols and curvature tensors.  This is unlike our previous formulas for $R^j$, such as \eq{FG}. 

\textit{But, of course, \eq{FixedCherubini} is incorrect in general:} for example, a simple line element for which it fails is:
\begin{equation}
 ds^2 = g_{00} (z) dt^2 + g_{11} (t) dx^2 +  g_{22}(y) dy^2  +  2dy dz + g_{33}(t) dz^2,
 \label{peculiar}
\end{equation} 
where we made explicit the coordinate dependence of the components.  
One can also use a closely related form of the metric to illustrate the fact that 
\eq{FixedCherubini} is not generally covariant and whether it holds or not depends on the coordinate system used. (This is to be expected because $Q^a$ is not a vector density.) Consider the following metric, obtained from \eq{peculiar} by setting $g_{22}=1$:
\begin{equation}
 ds^2 = g_{00} (z) dt^2 + g_{11} (t) dx^2 +   dy^2  +  2dy dz + g_{33}(t) dz^2.
 \label{peculiar1}
\end{equation}
One finds that \eq{FixedCherubini} holds.  But if we introduce a new coordinate $\zeta$ by $y=(1/2)\zeta^2$, so that the metric becomes:
\begin{equation}
 ds^2 = g_{00} (z) dt^2 + g_{11} (t) dx^2 +   \zeta^2 d\zeta^2  +  2\zeta d\zeta  dz + g_{33}(t) dz^2,
 \label{peculiar2}
\end{equation}
then we find that \eq{FixedCherubini} does \textit{not} hold! For the metrics in \eq{peculiar1}
or \eq{peculiar2}, the $L_{GB}$ actually vanishes so we are looking for a $Q^j$ with $\partial_jQ^j=0$. For the metric in  \eq{peculiar1}, we find that $Q^j=0$ for all $j$. But for the metric in \eq{peculiar2} we find
\begin{equation}
\bar{Q}^0 = \frac{-1}{2} \frac{g_{11,0}g_{00,3}}{\zeta \sqrt{g_{33}-1} g_{00}^{3/2} \sqrt{g_{11}}}
\end{equation}
which has $\partial_0\bar Q^0\neq 0$. Obviously, the expression for $Q^j$ is very much coordinate dependent. In any case, we do not expect an expression like \eq{FixedCherubini} to be valid because it arose through an incorrect mathematical operation.

One of the authors of ref. \cite{precherubini} (in which \eq{FixedCherubini} was first derived) has confirmed (F.W.Hehl, private communication) that
\eq{FixedCherubini} is indeed incorrect and the error arose due to an invalid identification of frame indices with spacetime indices in \cite{precherubini}. Thus, there is no simple expression for $R^j$ in the case of $D=4$ for a general metric (though, as we shall show later, one can find such an expression if the metric possesses a Killing vector; see Sec. \ref{sec:killing}).

\section{General Case: \LL\ Lagrangians in $D=2m$} \label{sec:2mDLL}

We shall now consider the general case of \LL\ Lagrangians in the critical dimension $D=2m$.
There exists a simple geometrical interpretation for the \LL\ Lagrangians being total derivatives of expressions involving tetrads.   Indeed, $\mathcal{L}_m$ is simply the Euler density of a $2m$-dimensional manifold, as can be seen, for example, in the section on index theorems in ref. \cite{Eguchi1980213}.  When working in $2m$ dimensions, $\mathcal{L}_m$ is a topologically invariant scalar density because it is a closed form and therefore locally exact \cite{Bidal}.  The Chern-Simons form $Q_m$ is defined to embody this quality: evaluating it between two connections $\Omega$ and $\Omega'$, one obtains $\mathcal{L}_m(\Omega) - \mathcal{L}_m(\Omega') = dQ_m(\Omega,\Omega')$. 
In particular, if we restrict ourselves to a local coordinate patch, then we can introduce  a flat connection $\Omega'=0$ and can therefore write $\mathcal{L}_m = dQ_m(\Omega,0)$.  Hence, the existence of the Chern-Simons form is by itself a proof that $\mathcal{L}_m$ is an exact differential.  Moreover, this means that $R^j$ can be thought of as providing the local coordinate description of the Chern-Simons form.   In Appendix \ref{chern}, we show how $\mathcal{L}_m$ can be locally rewritten as $\mathcal{L}_m(\Omega) = dQ(\Omega,0)$:
\begin{widetext}
\myeq{totalder}{
\mathcal{L}_m &= \left(d \Omega^{A_1 B_1} + \Omega^{A_1}_{~S_1} \wedge \Omega^{S_1 B_1} \right) \wedge \cdots \wedge 
\left(d \Omega^{A_m B_m} + \Omega^{A_m}_{~S_m} \wedge \Omega^{S_m B_m} \right) \wedge \epsilon_{A_1 B_1 \ldots A_m B_m} \\
&= m d \int_0^1 dt \Omega^{a_1 b_1} \wedge \Theta_t^{a_2 b_2} \wedge \cdots \wedge \Theta_t^{a_m b_m} \wedge \epsilon_{a_1 b_1 \cdots a_m b_m}.
}
\end{widetext}
where $\Theta_t = t d \Omega_t + t^2 \Omega \wedge \Omega$ is the curvature corresponding to the interpolated connection $\Omega_t = t \Omega$ which interpolates between $\Omega$ and $\Omega'=0$ as $t$ varies between $0$ and $1$.
Although these concepts are somewhat abstract, there exist many simple cases where it is actually quite easy to see how these Lagrangians can be rewritten as a total derivative.  In this sense, it is worthwhile to emphasize certain algebraic aspects of \LL\  Lagrangians, which we will now attempt.

To begin with, because all the indices are contracted with the antisymmetric $\epsilon$ tensor, and because the spin connections are antisymmetric when working in an orthonormal basis, we have  significant freedom in rearranging the terms and renaming indices.  Hence it is possible to perform a number of calculations without having to explicitly worry about the positioning of the indices; Appendix \ref{chern} is a good example of how advantageous this can be.

Second, because both $d \Omega$ and $\Omega \wedge \Omega$ are two-forms, they commute with each other.  Combined with the above fact regarding the freedom in renaming indices appropriately, this means that we  are able to use the binomial theorem to rewrite the Lagrangian as, figuratively, 
\begin{equation}
\mathcal{L}_m = (d \Omega + \Omega^2)^m = \sum \myc{m}{p} (d \Omega)^p (\Omega^2)^{m-p}.
\end{equation}   
The $p=0$ term in this expansion vanishes identically by antisymmetry. (For example, consider  the $m=1$ case, where the $p=0$ term would be
\myeq{antisymVanishes}{
\Omega^A_{~S} \wedge \Omega^{SB} \wedge \epsilon_{AB} = 2 \Omega^0_{~S} \wedge \Omega^{S1} \wedge \epsilon_{01} =0,
}
where, in the last equality, we used the antisymmetry of $\Omega$ and the fact that we are working in  two dimensions.)  Therefore, every term in the binomial expansion will be of the form $(d \Omega)^p (\Omega^2)^{m-p}$ with $p \geq 1$.  

Third, in $2m$ dimensions, $D' \epsilon_{a_1 \ldots a_{2m}}$ vanishes, where $D'$ is the covariant derivative corresponding to any connection $\Omega'$.  Indeed, denoting $D_g$ as the covariant derivative built through the metric connection $\Omega_g$, we have $D_g \epsilon=0$.  This means that 
\myeq{Depsilon}{
d \epsilon_{a_1 \ldots a_{2m}} &= \Omega^s_{g a_1} \epsilon_{s a_2 \ldots a_{2m}} + \cdots + \Omega^s_{g a_{2m}} \epsilon_{a_1  \ldots a_{2m-1} a_{s}} \\
&=0,
}
which vanishes because we require all of $\left\{ s,a_1,\ldots,a_{2m} \right\}$ to be different, which is impossible in $2m$ dimensions.  This then implies $D' \epsilon=0$.

Using these ideas, it is straightforward to rewrite some \LL\ Lagrangians as total derivatives directly, using the chain rule instead of abstract ideas involving Chern-Simons forms.  The $m=2$ case is demonstrated in Appendix \ref{cherubini}.  Since those equations appear, at first glance, challenging, we can here demonstrate the idea on an easier example to show that it is actually quite simple:
\myeq{easyExample}{
d\Omega^{ab}& \wedge \Omega^{cd} \wedge \epsilon_{abcd} \\
&= d(\Omega^{ab} \wedge \Omega^{cd}) \wedge \epsilon_{abcd} - \Omega^{ab} \wedge d \Omega^{cd} \wedge \epsilon_{abcd} \\
&= d(\Omega^{ab} \wedge \Omega^{cd}) \wedge \epsilon_{abcd} - \Omega^{cd} \wedge d \Omega^{ab} \wedge \epsilon_{abcd} \\
&= \frac{1}{2}d(\Omega^{ab} \wedge \Omega^{cd}) \wedge \epsilon_{abcd} \\
&= \frac{1}{2}d(\Omega^{ab} \wedge \Omega^{cd} \wedge \epsilon_{abcd})
}
where in the second-to-last step we equated the second-to-last line with the first, while in the last, we used $d \epsilon_{abcd}=0$.

It is therefore clear that one can express $\mathcal{L}_m$ as an exact differential in the form language.  As in the two-dimensional case, it is straightforward to use this equation to calculate the set of (non-unique) functions $R^j$ such that $L_m \sqrt{-g}= \partial_j R^j$.  One simply has to define an orthonormal basis $\omega^a$, use it to calculate the spin connections $\Omega^{ab}$, and use those in \eq{totalder}.  Therefore, $L_m \sqrt{-g}$ can easily be written as a total derivative of a set of functions of the metric and its derivatives in any critical dimension.

\subsection{\label{sec:killing} Spacetimes with a Killing Vector}
If we restrict ourselves to spacetimes which have a Killing vector, then $R^j$  takes a dramatically simpler form, which is worth emphasizing. This fact is a direct consequence of the results obtained in ref.\cite{Kolekar:2010dm} and the explicit proof is as follows:
Suppose the spacetime metric admits a timelike Killing  vector $\xi^a$ which we can choose, without loss of generality, to have the components: $\xi^a = (1,0,\cdots,0)$.
 Since $\mathcal{G}_{ab}=0$, in the critical dimension \cite{Kothawala:2009kc}, it follows from \eq{genEab} in the case of \LL\ models with $\nabla_aP^{abcd}=0$ that 
\begin{equation}
L_m=2P^{0cde} R_{0cde}
\end{equation} 
 We now note that
\begin{equation}
L_m=2P^{0cdb}R_{0cdb} = P^{0cdb}R_{acdb}\xi^a = 2P^{0cdb}\nabla_c \nabla_d \xi_b 
= \frac{1}{\sqrt{-g}}\partial_c \left( 2\sqrt{-g}P^{0cdb}\nabla_d \xi_b \right) 
\end{equation} 
where we used the fact that $P^{abcd}$ is divergence-free on every index and has the same symmetries as the Riemann tensor.  It follows that $L_m\sqrt{-g}=\partial_c R^c$ 
with $R^c=2\sqrt{-g}P^{0cdb}\nabla_d \xi_b$. Using the relations
$\nabla_d \xi_b = \Gamma_{b0d}$
and  find that  our final result can be written in the form: $L_m\sqrt{-g} = \partial_j R^j$ with 
\begin{equation} \label{KillingFG} R^j = -2 \sqrt{-g} P_a^{\phantom{a}b j 0} \Gamma_{0 b}^a. \end{equation}
In this case $R^j$ has the interpretation of being a component of the Noether potential corresponding to diffeomorphism invariance (see \cite{Kolekar:2010dm} for a discussion of Noether current in the context of action principles.).  This is interesting because it gives a direct physical interpretation to the local version of the Chern-Simons form in spacetimes which have symmetries. 

Obviously, similar results hold for spacetimes which have a spacelike Killing vector; if the Killing vector is taken to be, say, $\xi^j=\delta^j_1$ in a coordinate system, we can obtain a similar expression starting from the identity $\mathcal{G}^1_1=0$.
When the spacetime has more than one Killing vector, we  obtain, in general, more than one choice of $R^j$ which is again an example of the non-uniqueness pointed out before. The metric in \eq{simplemetric}, for example, has three obvious Killing vectors and hence one can write  simple expressions for $R^j$. They will differ from the expression in \eq{R0GB}, as well from each other by functions $f^j$ which satisfy $\partial_jf^j=0$.
Thus, for a large family of spacetimes, we have a very simple way to express the \LL\ Lagrangians as total derivatives.

\section{Conclusions}

We were led to this investigation and writing up of this paper in a fairly detailed, pedagogical, style because of significantly different perceptions of this issue amongst researchers. Those who are trained in the differential form language and approach general relativity from that perspective find it a rather trivial result that the action for \LL\ models is a surface term in the critical dimension. Given the fact that the action is an integral over an exact form (see \eq{totalder}), they find it rather strange that there should be any controversy in this matter! On the other hand, those approaching the subject from more traditional point of view of tensors, metric, Christoffel symbols, etc. do not find it so obvious that $L\sqrt{-g}=\partial_j R^j$ should hold in the critical dimension. This situation is made more confusing because we cannot build a vector density $R^j$ from the metric and its first derivatives. Further, no explicit expressions for $R^j$ seems to be given in the literature (except a couple of incorrect ones which we have discussed), contrary to the often expressed hope by our colleagues that such expressions must exist in the literature! All these prompted us to present our results in a rather detailed format, making the paper self-contained.

In order to clarify several contradictory comments in the literature, we have given a complete description of how the $2m$-dimensional \LL\ Lagrangians, and especially the two-dimensional Einstein-Hilbert Lagrangian, can be written locally as total derivatives $L_m \sqrt{-g} = \partial_j R^j$ with $R^j$ expressible in terms \textit{of the metric and its derivatives}. This $R^j$ has been calculated explicitly, as a function of the metric and its derivatives, for the two-dimensional Einstein-Hilbert case and we have given a constructive algorithm for the general case. We have also shown that $R^j$ can be thought of as the local version of the Chern-Simons form written in coordinate language and  is related to the Noether current in spacetimes which have a Killing vector.   

\begin{acknowledgments}
Large parts of this work were done with the support of the Perimeter Scholars International program at the Perimeter Institute.  A. Yale is supported by the Natural Sciences and Engineering Research Council of Canada. We thank Kinchal Banerjee, Dawood Kothawala, Rob Myers and Aseem Paranjape for several useful discussions and comments on the manuscript.  We further thank F. W. Hehl and C. Cherubini for useful email correspondence related to the material in Sec.\ref {sec:4DGB}, S.Deser and R.Jackiw for useful email correspondence related to the material in Sec. \ref{sec:2DEH}.
\end{acknowledgments}

\appendix
\section{Review of the Cartan Formalism} \label{sec:CartanReview}
In order to make sure that the equations are simple to follow, and also to establish notation, we shall provide a quick summary of the Cartan formalism, based on the language of differential forms, which is used extensively throughout this article.  This material is available in a number of other sources, such as \cite{gravitation}.

We have a $D$-dimensional manifold ${\cal M}$ with boundary $\partial {\cal M}$ on which we define a set of coordinates 
$\left\{ x^j \right\}$.  Since we will want to introduce a covariant derivative, we attach, locally on every coordinate patch in this manifold,  a cotangent bundle $T^*$ with basis vectors $\omega^a$.  These basis covectors are related to the coordinate basis $\left\{ dx^j \right\}$ by a set of functions $\omega^a_j$: $\omega^a = \omega^a_j dx^j$.  Next, we define the spin connection one-form $\Omega^a_{~b}$ through the first Cartan structural equation:
\begin{equation} \label{cartan1} d \omega^a + \Omega^a_{~b} \wedge \omega^b = 0, \end{equation}
for a torsion-free derivative.  In particular, for a coordinate basis (that is, $\omega^a = \delta^a_i dx^i$), the spin connection $\Omega^a_{~b} = \Gamma^a_{b j} dx^j$ is simply made of the Christoffel symbols.  This spin connection lets us define a covariant derivative $D$, whose action on $(p,q)$ tensors is given by
\myeq{defD}{
D &V^{a_1 \cdots a_p}_{b_1 \cdots b_q} = d V ^{a_1 \cdots a_p}_{b_1 \cdots b_q} \\
&+ \Omega^{a_1}_{~c} \wedge  V^{c a_2 \cdots a_p}_{b_1 \cdots b_q} + \cdots + \Omega^{a_p}_{~c} \wedge  V^{a_1 \cdots a_{p-1} c}_{b_1 \cdots b_q}\\
&- \Omega^c_{~b_1} \wedge V^{a_1 \cdots a_p}_{c b_2  \cdots b_q} - \cdots - \Omega^c_{~b_q} \wedge V^{a_1 \cdots a_p}_{b_1  \cdots b_{q-1} c}. 
}
The curvature two-form is then given by the second Cartan structural equation:
\begin{equation} \Theta^a_{~b} = d \Omega^a_{~b} + \Omega^a_{~c} \wedge \Omega^c_{~b} = \frac{1}{2}R^a_{~bcd} \omega^c \wedge \omega^d, \end{equation}
where $R^a_{bcd}$ is the Riemann tensor in the $\omega^a$ basis.  The Einstein-Hilbert Lagrangian may then be written
\myeq{EH}{ 
\Theta_{ab} \wedge * (\omega^a \wedge \omega^b) 
&= \frac{1}{2(D-2)!}R^{ab}_{mn}\omega^i_a\omega^j_b \varepsilon_{ijn_1 \cdots n_{D-2}} \\
&~~~~dx^m \wedge dx^n \wedge dx^{n_1} \wedge \cdots \wedge dx^{n_{D-2}} \\
&= R^{ji}_{mn}\delta^{mn}_{ij} \sqrt{-g} d^D x \\
&= R \sqrt{-g} d^D x ,
}
where $\varepsilon_{a_1 \cdots a_D}$ is the Levi-Civita symbol.

One can choose the basis for the cotangent bundle such that it is orthonormal: it is defined as
\begin{equation} \label{vielbeindef} g_{ij} = \eta_{ab} \omega^a_i \omega^b_j .\end{equation}
This formalism encodes all the spacetime information into the functions $\omega^a_i$, called vielbeins, while the Minkowski metric $\eta_{ab}$ is used to raise and lower indices.  In particular, if we demand that $D \eta_{ab}=0$, we will find that $\Omega_{ab}$ must be antisymmetric.  This means that there will be $\frac{1}{2}D(D-1)$ independent spin connections.  Therefore, for the two-dimensional case which interested us in Section \ref{sec:2DEH}, we have only one spin connection: $\Omega_{01}$.  

If we define a set of functions $\mu_{bc}^{~~a}$ by
\begin{equation} \label{mu} d \omega^a = -\frac{1}{2} \mu_{bc}^{~~a} \omega^b \wedge \omega^c, \end{equation}
Then it is easy to see that we can solve \eq{cartan1} by choosing
\begin{equation} \label{Omega} \Omega_{ab} = \frac{1}{2} \left( \mu_{abc} + \mu_{acb} - \mu_{bca} \right) \omega^c. \end{equation}
Indeed,
\begin{equation} \begin{split}
\Omega^a_{~b} \wedge \omega^b &= \frac{1}{2} \left( \mu^{a}_{bc} + \mu^{a}_{cb} - \mu_{bc}^{~~a} \right) \omega^c \wedge \omega^b \\
&= \frac{1}{2} \mu_{bc}^{~~a} \omega^b \wedge \omega^c \\
&= -d \omega^a .
\end{split} \end{equation}

\section{\label{app:der1} Details of Section \ref{der1}}
Let's begin by deriving \eq{R=dOmega} in detail, using an orthonormal basis such that $\Omega_{AB}$ is antisymmetric:
\myeq{proof}{
R \sqrt{-g} d^2x &= \Theta^{ab} \wedge *(\omega_a \wedge \omega_b) \\
&= \left( d \Omega^{ab} + \Omega^a_{~c} \wedge \Omega^{cb} \right) \wedge *(\omega_a \wedge \omega_b) \\
&= 2\left( d \Omega^{01} + \Omega^0_{~c} \wedge \Omega^{c1} \right) \wedge *(\omega_0 \wedge \omega_1) \\
&= 2d \Omega^{01} \wedge *(\omega_0 \wedge \omega_1) \\
&= 2d \left( \Omega^{01} \wedge *(\omega_0 \wedge \omega_1) \right) \\
&~~~~~+ \Omega^{01} \wedge \Omega^{~s}_0 \wedge *(\omega_s \wedge \omega_1) \\
&~~~~~+ \Omega^{01} \wedge \Omega_{1 s} \wedge *(\omega_0 \wedge \omega^s) \\
&= 2d \left( \Omega^{01} \wedge *(\omega_0 \wedge \omega_1) \right) \\
&= d \Omega_{01} d^2x.
}
Then, given an arbitrary two-dimensional metric $g_{j i}$, one can define an orthonormal basis for the cotangent bundle by choosing the gauge where $\omega_0^1=0$.  The functions $\omega^a_i$ which solve $g_{ij} = \omega^a_i \omega^b_j \eta_{ab}$ are then
\myeq{def1}{
\omega_0^0 &= \sqrt{-g_{00}} 							   \hspace{1.cm} \omega_0^1 = 0 \\
\omega_1^0 &= \frac{-g_{01}}{\sqrt{-g_{00}}} \hspace{1.cm} \omega_1^1 = \sqrt{ g_{11} - \frac{g_{01}^2}{g_{00}}} ,
}
such that we may write relations between the orthonormal basis $\omega^i$ and the coordinate basis $dx^i$:
\myeq{Relation_Omega_dx}{
\omega^0 &= \omega_0^0 dx^0 + \omega^0_1 dx^1  \hspace{2.cm}  \omega^1 = \omega^1_1 dx^1 \\
dx^0 &= \frac{1}{\omega^0_0} \left[ \omega^0 - \frac{\omega^0_1}{\omega^0_0} \omega^1 \right]  \hspace{1.5cm} dx^1 = \frac{1}{\omega^1_1} \omega^1. 
}
Having found our orthonormal basis, we can move on to calculate the quantities $\mu_{bc}^{~~a}$ defined by \eq{mu}, which then let us calculate the spin connection $\Omega_{01}$ defined by \eq{Omega}.  Differentiating $\omega^i$ gives us
\myeq{d omega}{
d \omega^0 &= \frac{ \omega^0_{1,0} - \omega^0_{0,1}}{\omega^0_0 \omega^1_1} \omega^0 \wedge \omega^1 \\
d \omega^1 &= \frac{\omega^1_{1,0}}{\omega^0_0 \omega^1_1} \omega^0 \wedge \omega^1.
}
This means that we have
\myeq{mu2}{
\mu_{100} &= -2 \frac{ \omega^0_{1,0} - \omega^0_{0,1}}{\omega_0^0 \omega^1_1} \\
\mu_{011} &= -2 \frac{ \omega^1_{1,0}}{\omega^0_0 \omega^1_1}
}
which, we must remember, are antisymmetric in the first two indices.  Note also that $\mu_{abc} = \mu_{ab}^{~~s} \eta_{s c}$.  Using now \eq{Omega} and differentiating it, we get
\myeq{dOmega}{
\Omega_{01} &= -\mu_{100} \omega^0_0 dx^0 + (\mu_{011} \omega^1_1 - \mu_{100} \omega^0_1) dx^1 \\
d \Omega_{01} &= \left[ (\mu_{100} \omega^0_0)_{,1} + (\mu_{011}\omega^1_1 - \mu_{100}\omega^0_1)_{,0} \right] dx^0 \wedge dx^1. 
}
Then, because the Einstein-Hilbert integrand can be written as in \eq{proof}, we can write
\myeq{split2}{
R \sqrt{-g} = \partial_0 R^0_{1} + \partial_1 R^1_{1},
}
where
\myeq{fg1}{
R^0_1 &= \mu_{100} \omega^0_1 - \mu_{011} \omega^1_1 \\
			&= \frac{1}{\sqrt{-g}} \left[ - \frac{g_{01}}{g_{00}} g_{00,1} + g_{11,0} \right] \\
R^1_1 &= \mu_{100} \omega^0_0 \\
			&= \frac{1}{\sqrt{-g}} \left[ \frac{g_{01}}{g_{00}}g_{00,0} - 2g_{01,0} + g_{00,1} \right].
}
This solution is clearly not symmetric in $t$ and $x$.  This is due to our gauge choice $\omega_0^1=0$; had we instead chosen the gauge $\omega_1^0=0$, we would have found
\myeq{fg2}{
R^0_2 &= \frac{1}{\sqrt{-g}} \left[ \frac{g_{01}}{g_{11}}g_{11,1} - 2g_{01,1} + g_{11,0} \right] \\
R^1_2 &= \frac{1}{\sqrt{-g}} \left[ - \frac{g_{01}}{g_{11}} g_{11,0} + g_{00,1} \right].
}
Because $R \sqrt{-g} = \partial_j R^j$ is linear, the most general solution can be written as the linear combination of $R_1$ and $R_2$:
\myeq{GeneralRmu}{
R \sqrt{-g} = \lambda \partial_j R_1^j + (1-\lambda)\partial_j R_2^j,
}
for a real parameter $\lambda$, which we can think of as embodying the extra gauge freedom we have in specifying vielbeins.  The final form of $R^j$ is then \eq{FG}.

\section{\label{app:der2} Conformal Transformation Method}
Knowing how $R$ and $\sqrt{-g}$ transform, and using $R \sqrt{-g} = \partial_j R^j$, we can discover how $R^j$ transforms under a conformal transformation $g_{j i} \rightarrow \bar{g}_{j i} = \Omega^2 g_{j i}$.  Indeed, we have
\myeq{RgTransformationLaw}{
\bar{R} \sqrt{-\bar{g}} 
&= \left[ R + \frac{2}{\Omega} g^{ef} \nabla_e \nabla_f \Omega - \frac{2}{\Omega^2} g^{ef} \nabla_e \Omega \nabla_f \Omega \right] \sqrt{-g} \\
&= \left[ R + 2 \nabla_e \left( g^{ef} \frac{1}{\Omega} \nabla_f \Omega \right) \right] \sqrt{-g} \\
&= R \sqrt{-g} + 2\sqrt{-g} \nabla_e \left( g^{ef} \frac{1}{\Omega} \nabla_f \Omega \right)   \\
&= R \sqrt{-g} + 2\sqrt{-g} \partial_e \left( \frac{1}{\Omega} g^{ei} \Omega_{,i} \right)  + 2\Gamma_{ab}^a \frac{1}{\Omega} g^{bi} \sqrt{-g} \Omega_{,i} \\
&= R \sqrt{-g} + 2\partial_e \left( \frac{\sqrt{-g}}{\Omega} g^{ei} \Omega_{,i} \right)   \\
&= \partial_j \left( R^j + \frac{2}{\Omega} \sqrt{-g} g^{j i} \Omega_{,i} \right).
}
Hence, $R^j$ transform as $R^j \rightarrow \bar{R}^j = R^j + \frac{2}{\Omega} \sqrt{-g} g^{j i} \Omega_{,i}$.  

n a diagonal gauge with $g_{01}=0$, it is straightforward to write
\begin{equation} R \sqrt{-g} = \partial_0 \left( \frac{g_{11,0}}{\sqrt{-g}} \right) + \partial_1 \left( \frac{g_{00,1}}{\sqrt{-g}} \right). \end{equation}
This suggests, for any gauge, an ansatz of the form:
\myeq{RmuAnsatz}{
R^j = \frac{1}{\sqrt{-g}}(& \lambda_1^j g_{00,0} + \lambda_2^j g_{01,0} + \lambda_3^j g_{11,0} \\
														&+ \lambda_4^j g_{00,1} + \lambda_5^j g_{01,1} + \lambda_6^j g_{11,1} ). 
}
The first solution, \eq{fg1} of section \ref{app:der1}, was found by setting $\omega^1_0=0$, which specified our gauge in the tangent space.  Similarly, we will impose a condition on $\lambda_i^j$:  that it either be a constant, or be a multiple of $\frac{g_{01}}{g_{00}}$; note that setting $\omega^1_0=0$ had also assumed $g_{00} \neq 0$.  This appears somewhat arbitrary, and it is, but we are here trying to guess the answer and this just so happens to work.  Finally, we demand that $R^j$ have the correct conformal transformation properties, as derived in earlier.  This yields
\myeq{Rmu1_conf}{
R^0 &= \frac{1}{\sqrt{-g}} \left( g_{11,0} - A\frac{g_{01}}{g_{00}}g_{00,1} + (A-1)g_{01,1} \right) \\
R^1 &= \frac{1}{\sqrt{-g}} \left( g_{00,1} - B\frac{g_{01}}{g_{00}}g_{00,0} + (B-1)g_{01,0} \right).
}
 If we substitute this into $(R \sqrt{-g} - \partial_j R^j)$, which we want to vanish, we will find some leftover terms such as $\left[2g_{00}^3 g_{01,01}g_{11} (A-B)\right]$, which have no other term against which they can cancel.  This forces $A=B$.  If we now use $A=B$ into our equation, we find $R \sqrt{-g} - \partial_j R^j =\frac{1-A}{2} [\ldots]$ where $[\ldots]$ is some function independent of $A$.  This sets $A=1$ and $B=-1$, such that we retrieve the old solution: \eq{fg1}.

To find the second solution, we perform the same steps, but instead of imposing that $\lambda_i^j$ either be a constant or a multiple of $\frac{g_{01}}{g_{00}}$, we now impose that $\lambda_i^j$ either be constant or a multiple of $\frac{g_{01}}{g_{11}}$.  After this time setting $A=-1$ and $B=1$, we get the second solution: \eq{fg2}.  As in Appendix \ref{der1}, we combine the solutions $R^j_1$ and $R^j_2$ into a general solution using \eq{GeneralRmu}.  This, once again, gives us \eq{FG} as desired.

\section{\label{cherubini} Derivation of equation $(\ref{works1})$}
\eq{works1} can be derived directly by simply rewriting the Lagrangian as a total derivative using the chain rule.  Since it is enlightening to see how this can be done without making use of the Chern-Simons form, we will provide the details of the calculations.    Working in an orthonormal frame, the spin connection $\Omega^{AB}$ will be antisymmetric and the Lagrangian can be written
\begin{widetext}
\myeq{GBLExpanded}{
\mathcal{L} &= \Theta^{ab} \wedge \Theta^{cd} \wedge \epsilon_{abcd} \\
&= \left[ d \Omega^{ab} \wedge d \Omega^{cd}
+ d\Omega^{ab} \wedge \Omega^c_{~f} \wedge \Omega^{fd}
+ \Omega^a_{~e} \wedge \Omega^{eb} \wedge d \Omega^{cd}
+ \Omega^a_{~e} \wedge \Omega^{eb} \wedge \Omega^c_{~f} \wedge \Omega^{fd} \right] \wedge \epsilon_{abcd} .
}
\end{widetext}
Since $d^2=0$, the first term may be written $d(\Omega^{AB} \wedge d \Omega^{CD})$.  The second term can be expanded as
\begin{widetext}
\myeq{onethird}{
&\sum_{a \cdots f} \left[ d  \Omega^{ab}  \wedge \Omega^c_{~f} \wedge \Omega^{fd} \right] \wedge \epsilon_{abcd}\\
&= \sum_{a \cdots f} \left[ d(\Omega^{ab} \wedge \Omega^c_{~f} \wedge \Omega^{fd}) + \Omega^{ab} \wedge d(\Omega^c_{~e} \wedge \Omega^{ed}) \right]  \wedge \epsilon_{abcd}\\
&= \sum_{a \cdots f} \left[ d(\Omega^{ab} \wedge \Omega^c_{~f} \wedge \Omega^{fd}) + \Omega^{ab} \wedge d \Omega^c_{~e} \wedge \Omega^{ed} - \Omega^{ab} \wedge \Omega^c_{~e} \wedge d \Omega^{ed} \right] \wedge \epsilon_{abcd}\\
&= \sum_{a \cdots f} \left[ d(\Omega^{ab} \wedge \Omega^c_{~f} \wedge \Omega^{fd}) + \Omega^{ab} \wedge d \Omega^c_{~a} \wedge \Omega^{ad} + \Omega^{ab} \wedge d \Omega^c_{~b} \wedge \Omega^{bd} - \Omega^{ab} \wedge \Omega^c_{~a} \wedge d \Omega^{ad} - \Omega^{ab} \wedge \Omega^c_{~b} \wedge d \Omega^{bd} \right] \wedge \epsilon_{abcd}\\
&= \sum_{a \cdots f} \left[ d(\Omega^{ab} \wedge \Omega^c_{~f} \wedge \Omega^{fd}) + \Omega^{ab} \wedge \left( \eta_{aa} d \Omega^{ca} \wedge \Omega^{ad} + \eta_{bb} d \Omega^{cb} \wedge \Omega^{bd} - \eta_{aa} \Omega^{ca} \wedge d \Omega^{ad} - \eta_{bb} \Omega^{cb} \wedge d \Omega^{bd} \right) \right] \wedge \epsilon_{abcd}\\
&= \sum_{a \cdots f} [ d(\Omega^{ab} \wedge \Omega^c_{~f} \wedge \Omega^{fd}) - \eta_{aa} \Omega^{ac} \wedge d \Omega^{ba} \wedge \Omega^{ad} - \eta_{bb} \Omega^{cb} \wedge d \Omega^{ab} \wedge \Omega^{bd} \\
& \hspace{3.63cm} + \eta_{aa} \Omega^{ad} \wedge \Omega^{ca} \wedge d \Omega^{ab} + \eta_{bb} \Omega^{db} \wedge \Omega^{cb} \wedge d \Omega^{ba} ] \wedge \epsilon_{abcd}\\
&= \sum_{a \cdots f} \left[ d(\Omega^{ab} \wedge \Omega^c_{~f} \wedge \Omega^{fd})  + d \Omega^{ab} \wedge \left( \eta_{aa} \Omega^{ac} \wedge \Omega^{ad} - \eta_{bb} \Omega^{cb} \wedge \Omega^{bd} + \eta_{aa} \Omega^{ad} \wedge \Omega^{ca} - \eta_{bb} \Omega^{db} \wedge \Omega^{cb} \right)\right] \wedge \epsilon_{abcd}\\
&= \sum_{a \cdots f} \left[ d(\Omega^{ab} \wedge \Omega^c_{~f} \wedge \Omega^{fd})  - d \Omega^{ab} \wedge \left( \Omega^c_{~a} \wedge \Omega^{ad} + \Omega^c_{~b} \wedge \Omega^{bd} + \Omega^c_{~a} \wedge \Omega^{ad} + \Omega^c_{~b} \wedge \Omega^{bd} \right) \right] \wedge \epsilon_{abcd}\\
&= \sum_{a \cdots f} \left[ d(\Omega^{ab} \wedge \Omega^c_{~f} \wedge \Omega^{fd}) - 2 d \Omega^{ab} \wedge \Omega^c_{~e} \wedge \Omega^{ed} \right] \wedge \epsilon_{abcd}\\
&= \sum_{a \cdots f} \left[ \frac{1}{3} d(\Omega^{ab} \wedge \Omega^c_{~f} \wedge \Omega^{fd}) \right] \wedge \epsilon_{abcd},
}
\end{widetext}
where we used the antisymmetry of $\epsilon_{abcd}$ to ensure that all of $a,b,c,d$ were different, and the antisymmetry of $\Omega_{ab}$ to ensure that $f$ could not be either $c$ or $d$.  The third term is equal to the second by antisymmetrizing the two pairs $(a \leftrightarrow c)$ and $(b \leftrightarrow d)$, while the fourth vanishes exactly by antisymmetry of $\Omega^{ab}$, as in \eq{antisymVanishes}.  Noting that the covariant derivative of $\epsilon$ must vanish, its exterior derivative will be of the form
\myeq{de}{ 
d \epsilon_{abcd} = \Omega^{~s}_a \epsilon_{sbcd} + \ldots + \Omega^{~s}_d \epsilon_{abcs}.
}
This is identically zero, as antisymmetry requires that $s$ be different than all of $a,b,c$ and $d$, which is impossible in four dimensions.  We can now easily write the Gauss-Bonnet Lagrangian as an exact differential:
\myeq{GB4}{
\mathcal{L} &= \Theta^{AB} \wedge \Theta^{CD} \wedge \epsilon_{ABCD} \\
&= \left( d \Omega^{AB} + \Omega^A_{~E} \wedge \Omega^{EB} \right) \wedge \left( d \Omega^{CD} + \Omega^C_{~F} \wedge \Omega^{FD} \right) \\
& ~~~~~\wedge \epsilon_{ABCD} \\
&= \left(d \Omega^{AB} \wedge d \Omega^{CD} + 2 d \Omega^{AB} \wedge \Omega^C_{~F} \wedge \Omega^{FD} \right) \wedge \epsilon_{ABCD} \\
&= d \left[ \left( \Omega^{AB} \wedge d \Omega^{CD} + \frac{2}{3} \Omega^{AB} \wedge \Omega^C_{~F} \wedge \Omega^{FD} \right) \wedge \epsilon_{ABCD} \right] \\
&= d \left[ \left( \Omega^{AB} \wedge \Theta^{CD} - \frac{1}{3} \Omega^{AB} \wedge \Omega^C_{~F} \wedge \Omega^{FD} \right) \wedge \epsilon_{ABCD} \right] .
}

\section{\label{chern} Proof that $\mathcal{L}_m = dQ$}
There exist many derivations of the Chern-Simons form $Q$; for an in-depth discussion see \cite{Nakahara, Bidal, Eguchi1980213}.  We will here give a slightly different (and more direct) derivation: we want to show that, locally, $\mathcal{L}_m = dQ$.  We introduce the interpolated connection $\Omega_t = t \Omega$, which interpolates between the connection $\Omega$ and the flat connection $\Omega'=0$, as well as its corresponding covariant derivative $D_t$ and curvature $\Theta_t$.  Then, we define the Chern-Simons form by
\myeq{CsDef}{
Q = m \int_0^1 dt \Omega^{a_1 b_1} \wedge \Theta_t^{a_2 b_2} \wedge \cdots \wedge \Theta_t^{a_m b_m} \wedge \epsilon_{a_1 b_1 \cdots a_m b_m} .
}
The Bianchi identity states that $D_t \Theta_t = 0$; a quick calculation shows that this is equivalent to the statement that $d (\Omega_1 \wedge \Omega_2) = d \Omega_1 \wedge \Omega_2 - \Omega_1 \wedge d \Omega_2$, and is therefore trivially satisfied:
\myeq{Bianchi}{
D_t \Theta_t^{ab} 
&= d(t^2 \Omega^a_{~s} \wedge \Omega^{sb}) + t \Omega^a_{~s} \wedge \Theta_t^{sb} + t \Omega^b_{~s} \wedge \Theta_t^{as} \\
&= t^2 d (\Omega^a_{~s} \wedge \Omega^{sb}) + t \Omega^a_{~s} \wedge (t d \Omega^{sb} + t^2 \Omega^s_{~t} \wedge \Omega^{tb}) \\
&~~~~+ t \Omega^b_{~s} \wedge (t d \Omega^{sa} + t^2 \Omega^s_{~t} \wedge \Omega^{ta})\\
&= t^2 \left[ d(\Omega^a_{~s} \wedge \Omega^{sb}) + \Omega^a_{~s} \wedge d \Omega^{sb} + \Omega^b_{~s} \wedge d \Omega^{as}\right] \\
&= 0,
}
where in the third equality we removed the terms symmetric in $(a \leftrightarrow b)$, knowing that we will always be contracting with $\epsilon_{a_1 b_1 \cdots a_m b_m}$.  Then, because the covariant derivative of a scalar is equal to its regular derivative, it is clear that
\myeq{dBecomesDt}{
dQ &= m d \left( \int_0^1 dt \Omega^{a_1 b_1} \wedge \Theta_t^{a_2 b_2} \wedge \cdots \wedge \Theta_t^{a_m b_m} \wedge \epsilon_{a_1 b_1 \cdots a_m b_m} \right) \\
&= m \int_0^1 dt D_t \Omega^{a_1 b_1} \wedge \Theta_t^{a_2 b_2} \wedge \cdots \wedge \Theta_t^{a_m b_m} \wedge \epsilon_{a_1 b_1 \cdots a_m b_m}.
}
Because of the antisymmetry on $\epsilon$, we have fair amount of freedom in reordering the indices in our equation.  In order to improve the readability of our equations, we shall therefore drop the indices entirely for this derivation.  (One needs to remember that $\Omega^{2m}=0$.) Then:
\begin{equation} \begin{split}
dQ=m \int_0^1 dt D_t \left( \Omega \Theta_t^{m-1} \right)
&= m \int_0^1 dt  (d \Omega + 2t \Omega^2)(t d\Omega + t^2 \Omega^2)^{m-1} \\
&= m \int_0^1 dt  t^{m-1} \sum_{p=0}^{m-1}(d \Omega + 2t \Omega^2) \myc{m-1}{p} (d \Omega)^p (t \Omega^2)^{m-1-p} \\
&= m \int_0^1 dt  t^{m-1} \sum_{p=0}^{m-1} \myc{m-1}{p} (d\Omega)^{p+1}t^{m-1-p} \Omega^{2(m-1-p)} \\
&~~~~~~~~~~+ 2mt^{m-1} \sum_{p=1}^{m-1} \myc{m-1}{p} (d\Omega)^pt^{m-p}\Omega^{2(m-p)} \\
&= m \int_0^1 dt  t^{m-1} \sum_{p=0}^{m-1} \myc{m-1}{p} (d\Omega)^{p+1}t^{m-1-p} \Omega^{2(m-1-p)} \\
&~~~~~~~~~~+ 2mt^{m-1} \sum_{p=0}^{m-2} \myc{m-1}{p+1} (d\Omega)^{p+1}t^{m-p-1}\Omega^{2(m-p-1)} \\
&= m\int_0^1 dt  t^{m-1} \sum_{p=0}^{m-1} (d \Omega)^{p+1} t^{m-1-p} \Omega^{2(m-p-1)} \left[ \myc{m-1}{p} + 2\myc{m-1}{p+1}\right] \\
&= m\int_0^1 dt  t^{m-1} \sum_{p=1}^{m} (d \Omega)^{p} t^{m-p} \Omega^{2(m-p)} \left[ \myc{m-1}{p-1} + 2\myc{m-1}{p}\right] \\
&= \sum_{p=1}^{m} \myc{m}{p} (d \Omega)^{p} \Omega^{2(m-p)} \\
&= \mathcal{L}_m,
\end{split} \end{equation}
where the second-to-last step involves some simple algebra:
\myeq{combinatorics}{
m& \int_0^1 dt t^{2m-p-1} \left[ \frac{(m-1)!}{(p-1)!(m-p)!} + 2 \frac{(m-1)!}{p!(m-1-p)!} \right] \\
&= \frac{1}{2m-p} \frac{m!}{(p-1)!(m-p-1)!} \left[ \frac{1}{m-p} + \frac{2}{p} \right] \\
&= \frac{1}{2m-p} \frac{m!}{p! (m-p)!} \left[ p + 2(m-p) \right] \\
&= \frac{m!}{p! (m-p)!}.
}


\begin{thebibliography}{99}
\bibitem{comment2} This curious result was first pointed out to one of us (T.P) by Dawood Kothawala and Kinjal Banerjee.
\end{thebibliography}
\end{document}